\documentclass[12pt,lettersize,journal,onecolumn]{IEEEtran}
\usepackage{subcaption} % For subfigure environment (recommended)
\usepackage{amsmath,amsfonts}
\usepackage{algorithmic}
\usepackage{algorithm}
\usepackage{array}
\usepackage[caption=false,font=normalsize,labelfont=sf,textfont=sf]{subfig}
\usepackage{textcomp}
\usepackage{stfloats}
\usepackage{url}
\usepackage{verbatim}
\usepackage{graphicx}
\usepackage{cite}
\usepackage{setspace}
\usepackage{longtable} 
\usepackage{makecell} % add this in the preamble
\usepackage{array}

\begin{document}

\title{Energy Efficient Point-to-Point PON-based\\[-0.9em] 
        Architecture for the Backhaul of a VLC \\[-0.9em] 
         System \vspace{-0.5em}}

\author{
    Wafaa~B.~M.~Fadlelmula,  
    Sanaa~Hamid~Mohamed,  
    Taisir~E.~H.~El-Gorashi,\\[-0.7em]  
    and Jaafar~M.~H.~Elmirghani%

}



\maketitle

\begin{abstract}
This paper proposes a point-to-point passive optical network (P2P-PON) architecture as an energy-efficient and low-latency backhaul solution for visible light communication (VLC)-enabled indoor fog computing systems. The proposed architecture passively interconnects VLC access points and distributed in-building fog servers through dedicated optical links, enabling flexible peer-to-peer connectivity and efficient traffic aggregation. A mixed integer linear programming (MILP) framework is developed to jointly optimize processing resource allocation, traffic routing, power consumption, and end-to-end queuing delay across a multi-layer fog computing infrastructure. The model explicitly captures the power consumption of both networking and processing elements and incorporates a piecewise linear approximation of an M/M/1 queuing model to represent delay-sensitive applications. The performance of the proposed P2P-PON architecture is evaluated and compared against an arrayed waveguide grating router (AWGR)-based PON architecture under multiple indoor traffic scenarios. The results show that the proposed P2P-PON architecture reduces total power consumption by up to 64\% under power-aware optimization and by 15\% under delay-aware optimization, while reducing average end-to-end queuing delay by up to 76\% compared to the AWGR-PON architecture, due to the improved in-building connectivity and more effective utilization of distributed fog resources.

\end{abstract}


\section{Introduction}
\IEEEPARstart{T}{he} exponential growth of mobile data traffic, driven by bandwidth-intensive applications such as augmented reality, virtual reality, and ultra-high-definition video streaming, has created unprecedented demands on wireless communication infrastructure. This surge, together with the widespread adoption of Internet of Things (IoT) devices and the increasing reliance on edge intelligence, requires communication systems that deliver high data rates and low latency while maintaining strict energy efficiency~\cite{horvath2020passive,wang2023edge,altowaijri2025synergistic}.

Visible Light Communication (VLC) has emerged as a promising complementary technology to traditional radio frequency (RF) systems. By leveraging the vast unlicensed visible light spectrum, VLC offers high spatial reuse, enhanced security due to light confinement, and immunity to electromagnetic interference~\cite{karunatilaka2015led}. These properties make VLC particularly attractive for dense indoor deployments, such as offices, campuses, and smart buildings. However, the performance and scalability of VLC systems are heavily influenced by the design of the underlying backhaul network interconnecting access points and higher-layer processing resources.

In parallel, fog computing has gained attention as an enabling paradigm for low-latency and context-aware services. Unlike cloud-centric architectures that rely on remote data centers, fog computing distributes processing and storage resources across multiple layers closer to end users \cite{das2023review}. Integrating VLC access networks with fog computing presents significant opportunities for energy-efficient indoor communication systems. This hierarchical approach reduces latency, alleviates core network congestion, and improves energy efficiency by localizing computation whenever possible. By distributing computational resources from user devices and room-level fog servers to building-level and campus-level fog nodes, tasks can be executed closer to user, which will reduce latency and power consumption. Nevertheless, the effectiveness of fog computing critically depends on the capacity, flexibility, and energy efficiency of the access networks that interconnect distributed processing nodes.

Passive Optical Networks (PONs) present a compelling candidates for indoor backhaul, offering inherent energy efficiency through passive optical components. While PON-based architectures have been extensively studied for access networks \cite{abbas2016next, arya2025historical, horvath2020passive}, their application to indoor fog computing environments with VLC integration remains largely unexplored. Existing PON solutions, such as Arrayed Waveguide Grating Router (AWGR)-based architectures, provide passive connectivity but may suffer from wavelength blocking and limited flexibility in dynamic fog environments \cite{fadlelmula2025energy}.

To overcome these limitations, this paper proposes a novel point-to-point (P2P) PON-based backhaul architecture tailored for VLC-enabled indoor fog computing systems. Unlike our previously proposed PON architecture, the proposed architecture eliminates wavelength routing constraints and provides direct passive P2P connectivity between access points, room fog servers, and the optical line terminal (OLT). This design enhances routing flexibility, enables more efficient workload placement, and reduces networking power consumption by avoiding the use of tunable optical components.

The main contributions of this work are:

\begin{itemize}
    \item A novel P2P PON-based backhaul architecture for VLC-enabled indoor fog computing systems that removes wavelength routing constraints and improves in-building connectivity.
    \item A detailed Mixed Integer Linear Programming (MILP) model that jointly optimizes processing placement, networking power consumption, and end-to-end queuing delay across a multi-layer fog architecture.
    \item Incorporation of a piecewise linear M/M/1 queuing delay model to capture delay-energy trade-offs.
    \item Extensive evaluation against an AWGR-based PON architecture, showing significant reductions in total power consumption and improved delay performance under various scenarios.
\end{itemize}

The remainder of the paper is organized as follows. Section~\ref{sec:related-work} reviews the related work. Section~\ref{sec:in-building} presents the system model and describes the proposed P2P-PON backhaul architecture. Section~\ref{sec:MILP} introduces the MILP formulations for joint power and delay optimization. Performance evaluation results are discussed in Section~\ref{sec:per_eval}, and conclusions with future research directions are provided in Section~\ref{sec:conc}.

\section{Related Work}
\label{sec:related-work}
This work intersects several research domains, including VLC access networks, fog computing architectures, PONs and energy-efficient resource allocation using optimization frameworks. This section reviews the relevant studies in these areas and highlights the research gap. 

VLC has attracted significant research interest due to its ability to provide high data rates while leveraging existing lighting infrastructure \cite{matheus2019visible,mapunda2020indoor}. Extensive surveys have addressed VLC fundamentals, applications, and challenges, as well as resource allocation and multi-user transmission techniques \cite{matheus2019visible,mapunda2020indoor,obeed2019optimizing}. These works primarily focus on the access layer and physical-layer optimization, with limited attention to backhaul integration and network-level architectural considerations.

Several studies have investigated wired backhauling solutions for VLC access points. Power line communication (PLC) has been proposed as a cost-effective approach that exploits existing electrical wiring. In \cite{song2015indoor}, PLC is used to backhaul VLC access points by coupling data signals onto power lines. Although PLC-based solutions reduce deployment cost, their limited achievable data rates constrain scalability for high-capacity indoor VLC systems. Consequently, fiber-based backhauling has been explored to support higher throughput and reliability.

Fiber-connected VLC architectures have been demonstrated using LAN and optical networking technologies. Experimental studies have shown the feasibility of integrating VLC access points with fiber backbones to support full-duplex transmission and multi-user access \cite{mark2014ethernet}. Wireless optical backhauling has been proposed as an alternative to wired solutions. In \cite{kazemi2018wireless}, a wireless optical backhaul was introduced for optical attocell networks to enhance coverage and reduce interference. While this approach improves flexibility, it primarily addresses communication aspects and does not consider joint optimization of networking and computing resources. Other works investigated hybrid VLC/RF or optical backhaul solutions, focusing on throughput and connectivity improvements \cite{aboagye2020vlc}. Overall, these studies emphasize physical-layer performance and do not address processing resource placement or integrated network–compute optimization.

Beyond VLC-specific backhaul studies, resource allocation and energy efficiency have been extensively investigated in fog and mobile edge computing (MEC) architectures. Prior work has shown that deploying computing resources closer to the network edge can reduce latency and energy consumption \cite{jazayeri2021latency,cheng2018energy}. Comprehensive surveys highlight the benefits of multi-tier fog architectures over centralized cloud solutions \cite{costa2022computational,vergara2023comprehensive}. Optimization frameworks, including MILP-based formulations, have been proposed to jointly allocate communication and computing resources under energy constraints and latency constraints \cite{aletri2020optimum, mahmoudi2023optimal}. However, most of these studies abstract the access network and do not explicitly model VLC-specific backhaul architectures or in-building optical connectivity.

Recent work has explored the integration of VLC with fog and MEC systems to support latency-sensitive and energy-efficient applications \cite{xue2022flexible}. Other studies examined energy and spectral-efficiency optimization in heterogeneous VLC-based networks \cite{aboagye2020vlc,yang2019learning}. Despite these advances, existing VLC-related works generally treat the backhaul, access network, and processing layers independently, without a unified optimization framework that jointly considers networking power, processing power, workload assignment, and delay performance.

AWGR-PON architectures have recently been investigated for supporting in-building VLC systems with integrated fog computing capabilities. In our earlier work \cite{fadlelmula2023energy},, we proposed an AWGR-PON-based indoor architecture and formulated an MILP model to optimize workload placement with the objective of minimizing total power consumption. The results showed that integrating multi-tier fog nodes within the building can significantly reduce both networking and processing power consumption compared to cloud-centric approaches. This work was further extended in \cite{fadlelmula2025energy}, where additional architectural refinements and a more comprehensive evaluation were presented. However, AWGR-PON-based connectivity relies on signal splitting to serve multiple VLC access points and processing nodes, which, in some of the studied scenarios, reduced the effective bandwidth available to processing resources. This affected traffic steering and workload placement between access points and in-building fog nodes. Moreover, the previous AWGR-PON-based studies focused on power consumption minimization and did not include end-to-end delay modeling.

In contrast, the P2P-PON architecture provides dedicated optical links between VLC access points and in-building fog nodes, eliminating bandwidth sharing due to signal splitting and enabling direct access to distributed processing resources. Building on this architecture, this paper presents a comprehensive modeling and optimization framework for an indoor VLC access network integrated with a multi-tier fog and cloud architecture. The proposed approach jointly optimizes networking and processing power consumption, workload assignment, and delay performance within a unified formulation, addressing key limitations identified in prior VLC backhaul network architecture. A comparative summary of the most relevant works and the contributions of this paper are provided in Table~I.

\begin{table*}[htbp]
\centering
\footnotesize
\caption{Comparison of Related Work and Contributions}
\label{tab:related_work_summary}
\begin{tabular}{|
p{2.2cm}|
p{1.5cm}|
p{2.5cm}|
p{2.2cm}|
p{2.2cm}|
p{2.2cm}|
p{2.2cm}|
}
\hline
\textbf{Reference} &
\textbf{VLC Access} &
\textbf{Optical / Wired Backhaul} &
\textbf{Fog / Edge Processing} &
\textbf{Cloud Processing} &
\textbf{Energy Optimization} &
\textbf{Delay Consideration} \\
\hline

Song \textit{et al.} \cite{song2015indoor} &
\checkmark & PLC & -- & -- & -- & -- \\
\hline

Philip \cite{mark2014ethernet} &
\checkmark & Fiber (LAN) & -- & -- & -- & -- \\
\hline

Kazemi \textit{et al.} \cite{kazemi2018wireless} &
\checkmark & Optical Wireless & -- & -- & -- & \checkmark \\
\hline

Aboagye \textit{et al.} \cite{aboagye2020vlc} &
\checkmark & Hybrid RF/VLC & -- & -- & \checkmark & \checkmark \\
\hline

Cheng \textit{et al.} \cite{cheng2018energy} &
-- & Wireless & \checkmark & -- & \checkmark & -- \\
\hline

Jazayeri \textit{et al.} \cite{jazayeri2021latency} &
-- & Wireless & \checkmark & -- & \checkmark & \checkmark \\
\hline

Yang \textit{et al.} \cite{yang2019learning} &
\checkmark & Hybrid RF/VLC & \checkmark & -- & \checkmark & \checkmark \\
\hline

Xue \textit{et al.} \cite{xue2022flexible} &
\checkmark & VLC Attocell & \checkmark & -- & -- & \checkmark \\
\hline

Fadlelmula \textit{et al.} \cite{fadlelmula2023energy} &
\checkmark & AWGR-PON & \checkmark & \checkmark & \checkmark & -- \\
\hline

\textbf{This work} &
\checkmark &
\textbf{P2P-PON} &
\checkmark &
\checkmark &
\checkmark &
\checkmark \\
\hline

\end{tabular}
\end{table*}

\section{The System Model}
\label{sec:in-building}

\begin{figure}[t]
\centering
\includegraphics{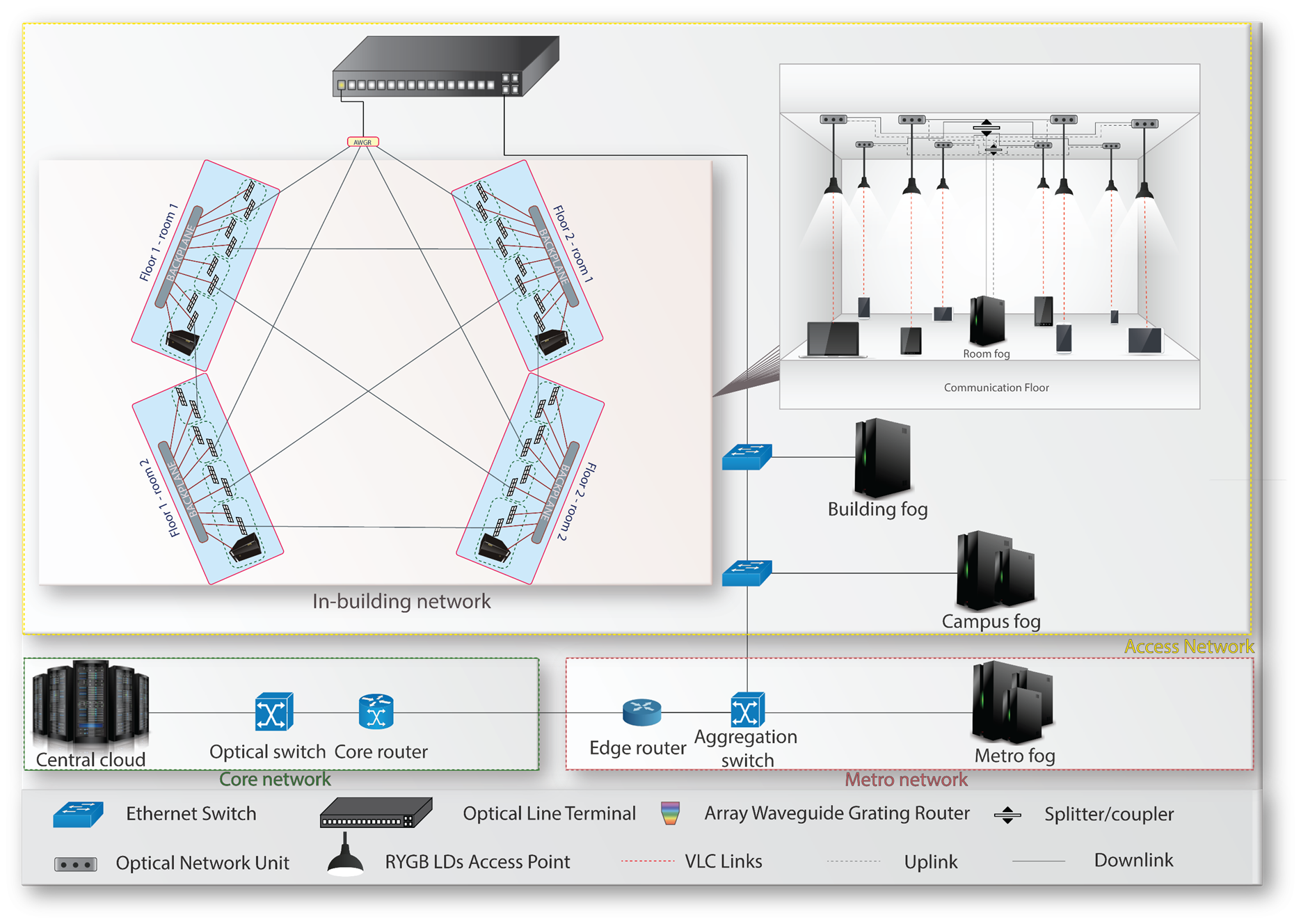}
\caption{The in-building S\&L VLC backhaul system.}
\label{figure1}
\end{figure}

The P2P-PON provides passive connectivity for the in-building backhaul network. As shown in Fig. \ref{figure1} The architecture is deployed within a multi-room building, where access points and room fog servers (RFSs) are interconnected through passive optical components.

Within each room, access points are directly connected to a passive polymer backplane, which enables local communication among access points and the RFS located in the same room. The access points and the RFS in each room are logically organized into multiple groups. The number of groups per room is determined by the total number of rooms in the building. For the considered building layout, each room is divided into four groups. Three groups consist of two access points each, while the fourth group includes two access points and the RFS.

In each group, one access point is designated to maintain a direct P2P optical connection with an access point located in a different room or with the OLT. This access point acts as a relay for intra-room traffic originating from other access points within the same group, in addition to transmitting its own traffic. As a result, traffic from access points without direct external connectivity can be forwarded through the designated relay access point.

The availability of routing paths in the architecture depends on the number of groups per room and the established inter-room P2P links. Each group maintains direct optical connections to either groups in other rooms or to the OLT. Indirect paths via intermediate access points are also supported and can be used to reach remote destinations when a direct link is not available. To illustrate the connectivity, consider the access points in the first room. One access point maintains a direct optical link to the OLT. Another access point is directly connected to an access point in the third room, while additional access points are connected to access points in the second and fourth rooms through dedicated P2P optical links. These connections collectively enable multiple routing options for traffic exchange across the building.

The main components of the PON-based backhaul network are described as follows.

\begin{itemize}
    \item \textbf{Optical Network Units (ONUs):}  
    Each access point and RFS is equipped with an optical network unit (ONU). Each ONU operates on a fixed, pre-assigned wavelength. 

    \item \textbf{Optical Line Terminal (OLT):}  
    The OLT aggregates upstream traffic from all rooms and forwards it to higher network layers. In the downstream direction, the OLT delivers traffic to the appropriate rooms based on the fixed wavelength assignment. The OLT performs traffic aggregation and network management functions.

    \item \textbf{Passive Polymer Backplane:}  
    A passive polymer backplane is deployed within each room to provide high-speed intra-room connectivity. 
    The backplane interconnects all access points and the RFS located in the same room. 

    \item \textbf{Arrayed Waveguide Grating (AWG):}  
    A passive arrayed waveguide grating (AWG) is used to aggregate upstream optical signals from multiple rooms into a single optical fiber connected to a port of the OLT. In the downstream direction, the AWG passively separates the wavelengths and forwards them to their corresponding rooms.
\end{itemize}

A multi-layer fog computing architecture is considered, where processing resources are distributed across several layers of the network, from the user devices to the core cloud.

\begin{itemize}
    \item \textbf{User Devices:}  
    User devices within each room generate computation requests and may also contribute idle processing resources by serving tasks originating from other devices within the building. A single type of mobile device is assumed to reduce system complexity. The processing capabilities of user devices are limited compared to dedicated fog servers.

    \item \textbf{Room Fog Servers (RFSs):}  
    RFSs are located within individual rooms and consist of personal computing devices such as laptops or desktops. Homogeneous laptops are assumed across all rooms.

    \item \textbf{Building Fog Servers (BFSs):}  
    BFSs provide higher processing capacities than RFSs and are implemented using dedicated servers. These servers serve computation requests aggregated from multiple rooms within the building.

    \item \textbf{Campus Fog Servers (CFSs):}  
    CFSs handle computation requests from multiple buildings within a campus. The CFS is located in the access network and is directly connected to the OLT via an Ethernet switch. A single server is assumed at this layer.

    \item \textbf{Metro Fog Servers (MFSs):}  
    MFSs are deployed in the metro network and offer high processing capacities. The metro network is modeled using a single edge router and a single aggregation switch connected to the metro fog server.

    \item \textbf{Central Cloud Servers (CCSs):}  
    CCSs are located in the core network and provide large-scale processing resources. For simplicity, a single cloud server is assumed for task execution.
\end{itemize}

\section{MILP Model for Joint Energy-Delay Optimization}
\label{sec:MILP}

In this section, we present the MILP model for our proposed P2P-PON architecture. We compare our approach against the AWGR-based wavelength-routed PON  architecture from our prior work \cite{fadlelmula2025energy}.

\subsection{Baseline: AWGR-based Wavelength-Routed PON}
\label{subsec:baseline}

Our baseline is the AWGR-based P2P-PON fog architecture presented in \cite{fadlelmula2025energy}. This architecture uses AWGRs to route traffic between nodes, requiring explicit wavelength assignment and imposing AWGR routing constraints.

The baseline model minimizes total power consumption while maximizing task acceptance:
\begin{equation}
\text{Minimize} \quad \alpha \, TPC - \tau \, \Upsilon
\end{equation}

where $TPC$ is total power consumption (processing + networking), $\Upsilon$ is the number of accepted tasks, and $\alpha$, $\tau$ 
are weight factors.

The complete mathematical formulation is available in \cite{fadlelmula2025energy}, Section IV. 

\subsection{Proposed Model: P2P-PON Architecture}
\label{subsec:proposed}

Our proposed architecture eliminates AWGRs and wavelength 
constraints, enabling direct peer-to-peer connectivity. This 
section presents the complete MILP formulation.

The sets, parameters, and variables of the MILP model for the PON-based backhaul architecture are defined as follows:
\begin{center}
\footnotesize
\begin{longtable}{l p{14cm}}
\caption{Notation Used in the MILP Model} \\
\hline
\textbf{Symbol} & \textbf{Description} \\
\hline
\multicolumn{2}{l}{\textbf{Sets}} \\
\hline
$N$ & Set of all nodes. \\
$N_m$ & Set of neighbors of node $m$, $m \in N$. \\
$UD$ & Set of user devices generating processing demands. \\
$PUD$ & Set of user devices providing processing resources. \\
$RF$ & Set of room fog servers (RFSs). \\
$P$ & Set of all processing nodes (user devices, RFS, BFS, CFS, MFS, cloud). \\
$AP$ & Set of access points connected to source user devices. \\
$CP$ & Set of access points connected to processing user devices. \\
$Q$ & Set of networking devices (OLT, switches, routers). \\
$PI$ & Set of linear segments used to approximate the nonlinear queuing delay function. \\
$MR$ & Set of access points connected to source user devices using red wavelength. \\
$MY$ & Set of access points connected to source user devices using yellow wavelength. \\
$MGB$ & Set of access points connected to source user devices using green/blue wavelength. \\
$DR$ & Set of access points connected to processing user devices using red wavelength. \\
$DY$ & Set of access points connected to processing user devices using yellow wavelength. \\
$DGB$ & Set of access points connected to processing user devices using green/blue wavelength. \\
$U$ & Set of additional processing-related access points. \\
\hline
\multicolumn{2}{l}{\textbf{Processing Parameters}} \\
\hline
$D_u$ & Processing demand of user device $u$, $u \in UD$. \\
$\Omega_d$ & Processing capacity of node $d$, $d \in P$. \\
$MP_d$ & Maximum power consumption of processing node $d$, $d \in P$. \\
$I_d$ & Idle power consumption of processing node $d$, $d \in P$. \\
$R_d$ & Power consumption per processing unit, $R_d = (MP_d - I_d)/\Omega_d$. \\
$\mu_d$ & Service rate of processing node $d$ (tasks/second), $d \in P$. \\
$\rho_d$ & Processing time per unit demand at node $d$, $d \in P$. \\
\hline
\multicolumn{2}{l}{\textbf{Networking Parameters}} \\
\hline
$R_u$ & Data rate demand of user device $u$, $u \in UD$. \\
$L_{mn}$ & Capacity of physical link $(m,n)$. \\
$DRR_u$ & Data rate ratio, $DRR_u = R_u / D_u$. \\
$\tau_{mn}$ & Transmission delay on link $(m,n)$. \\
$MQ_i$ & Maximum power consumption of networking device $i$, $i \in Q$. \\
$\gamma_i$ & Idle power consumption of networking device $i$, $i \in Q$. \\
$BQ_i$ & Bit rate capacity of networking device $i$, $i \in Q$. \\
$\phi_{q_i}$ & Power consumption per bit for device $i$, $\phi_{q_i} = (MQ_i - \gamma_i)/BQ_i$. \\
\hline
\multicolumn{2}{l}{\textbf{Access Point Parameters}} \\
\hline
$PR_{\max}$ & Maximum power consumption of access point using red wavelength. \\
$IR$ & Idle power consumption of access point. \\
$BP$ & Bit rate capacity of access point. \\
$\Lambda$ & Power per bit for access point, $\Lambda = (PR_{\max}-IR)/BP$. \\
$PY_{\max}$ & Maximum power consumption of access point using yellow wavelength. \\
$IY$ & Idle power consumption of access points using yellow wavelength. \\
$\Delta$ & Power per bit for access points using yellow wavelength, $\Delta = (PY_{\max}-IY)/BP$. \\
$PGB_{\max}$ & Maximum power consumption of access point using green/blue wavelength. \\
$IGB$ & Idle power consumption of access points using green/blue wavelength. \\
$\Pi$ & Power per bit for access points using green/blue wavelength, $\Pi = (PGB_{\max}-IGB)/BP$. \\
$PU_{\max}$ & Maximum power consumption of an ONU. \\
$\mathbb{U}$ & Idle power consumption of an ONU. \\
$BU$ & Bit rate of an ONU. \\
$\chi$ & Power per bit for an ONU, $\chi = (PU_{\max}-\mathbb{U})/BU$.\\
\hline
\multicolumn{2}{l}{\textbf{Queuing Delay Parameters}} \\
\hline
$RA_{mnp}$ & Slope of line for linear segment $p$ for queuing delay on link $(m,n)$, $m \in N, n \in N_m, p \in PI$. \\
$TI_{mnp}$ & Breakpoint of linear piece $p$ for linear approximation of delay on link $(m,n)$, $m \in N, n \in N_m, p \in PI$. \\
$UB_{mn}$ & Upper bound value of queuing delay on link $(m,n)$, $m \in N, n \in N_m$. \\
\hline
\multicolumn{2}{l}{\textbf{Optimization Parameters}} \\
\hline
$\alpha$ & Weight factor for energy consumption in objective function. \\
$\beta$ & Weight factor for delay in objective function. \\
$M$ & Large constant for queuing delay linearization. \\
$Z$ & Large positive constant for binary variable relationships. \\
\hline
\multicolumn{2}{l}{\textbf{Decision Variables}} \\
\hline
$\psi^{ud}$ & Processing demand assigned from user $u$ to processing node $d$. \\
$\xi^{ud}$ & Binary: 1 if user $u$ assigns task to processor $d$, 0 otherwise. \\
$\beta_d$ & Binary: 1 if processing node $d$ is activated, 0 otherwise. \\
$\Theta_i$ & Binary: 1 if networking device $i$ is activated, 0 otherwise. \\
$\lambda_{mn}^{ud}$ & Traffic flow from user $u$ to processor $d$ on link $(m,n)$. \\
$\lambda^{ud}$ & Total traffic flow from user $u$ to processor $d$. \\
$\lambda_i$ & Total traffic aggregated by networking device $i$. \\
$\mu_i$ & Total traffic generated by access point $i$. \\
$\sigma_i$ & Total traffic received by access point $i$. \\
$TR_{mn}$ & Total traffic on link $(m,n)$, $m \in N, n \in N_m$. \\
$TB_{mn}^{ud}$ & Binary: 1 if traffic from $(u,d)$ traverses link $(m,n)$, 0 otherwise. \\
$HB_{mn}$ & Binary: 1 if there is traffic on link $(m,n)$, 0 otherwise. \\
\hline
\multicolumn{2}{l}{\textbf{Delay Variables}} \\
\hline
$T^{ud}$ & End-to-end delay for task from user $u$ to processing node $d$. \\
$T_q^{ud}$ & Queuing delay for task from $u$ at processing node $d$. \\
$U_d$ & Utilization of processing node $d$. \\
$H_{mn}$ & Queuing delay on link $(m,n)$, $m \in N, n \in N_m$. \\
$X_{mn}^{ud}$ & Queuing delay for traffic from $u$ to $d$ traversing link $(m,n)$. \\
$LD_{ud}$ & Total queuing delay for traffic from source $u$ to destination $d$, $u \in UD, d \in P$. \\
$ED_u$ & End-to-end queuing delay experienced by user device $u$, $u \in UD$. \\

\hline
\multicolumn{2}{l}{\textbf{Objective Components}} \\
\hline
$TPC$ & Total power consumption. \\
$TD$ & Total delay (seconds). \\
$PC$ & Processing power consumption. \\
$PN$ & Networking power consumption. \\
$PAP$ & Power consumption of access points connected to source user devices. \\
$PCP$ & Power consumption of access points connected to processing user devices. \\
$AR$ & power consumption of access points connected to source user device using the red wavelength. \\ 
$AY$ & Power consumption of access points connected to source user device using the yellow wavelength. \\ 
$AGB$ & Power consumption of access points connected to source user device using the green/blue wavelength. \\
$PR$ & Power consumption of access points connected to the processing user device using the red wavelength.\\
$PY$ & Power consumption of access points connected to the processing user device using the yellow wavelength. \\
$PGB$ & Power consumption of access points connected to the processing user device using the green/blue wavelength.\\
$PONU$ & Power consumption of ONU attached to RFSs.\\
$PQ$ & Power consumption of networking devices including the OLT, Ethernet switches, aggregation switch, edge router and the optical switch. \\
\hline
\end{longtable}
\end{center}
\vspace{-3em}
The variables are related as follows: 
\vspace{-1.5em}
\begin{equation}
\mu_i = \sum_{u \in UD} \sum_{d \in P} \sum_{n \in N_i}
\left( \lambda_{in}^{ud} + \lambda_{ni}^{ud} \right),
\quad \forall i \in AP
\label{eq:ap_traffic}
\end{equation}
\vspace{-1.5em}
\begin{equation}
\sigma_i = \sum_{u \in UD} \sum_{d \in P} \sum_{n \in N_i}
\left( \lambda_{in}^{ud} + \lambda_{ni}^{ud} \right),
\quad \forall i \in CP
\label{eq:cp_traffic}
\end{equation}
\vspace{-1.5em}
\begin{equation}
\lambda_i = \sum_{u \in UD} \sum_{d \in P} \sum_{n \in N_i}
\left( \lambda_{in}^{ud} + \lambda_{ni}^{ud} \right),
\quad \forall i \in Q
\label{eq:network_traffic}
\end{equation}

\vspace{-0.5em}
Equations~\eqref{eq:ap_traffic}, \eqref{eq:cp_traffic}, and \eqref{eq:network_traffic} calculate the total traffic aggregated by different network devices. Equation~\eqref{eq:ap_traffic} computes the traffic handled by access points connected to source user devices, while Equation~\eqref{eq:cp_traffic} computes traffic for access points connected to processing user devices. Equation~\eqref{eq:network_traffic} aggregates traffic passing through networking devices such as the OLT, switches, and routers.

The objective is to minimize a weighted combination of total energy consumption and total end-to-end delay:
\vspace{-1em}
\begin{equation}
\text{Minimize} \quad \alpha \, \, TPC + \beta \, \, TD
\label{eq:objective_joint}
\end{equation}
\vspace{-2.5em}

where:
\vspace{-0.5em}
\begin{equation}
TPC = PC + PN \label{eq:total_power}    
\end{equation}
\vspace{-3em}

\textbf{Processing Power Consumption:}
\vspace{-1.3em}
\begin{equation}
PC = \sum_{u \in UD} \sum_{d \in P}\psi^{ud} R_d + \beta_d I_d
\label{eq:processing_power}
\end{equation}
\vspace{-1em}
\textbf{Networking Power Consumption:}
\vspace{-0.5em}
\begin{equation}
PN = PAP + PCP + AR + AY + AGB + PR + PY + PGB + PONU + PQ,
\label{eq:networking_power}
\end{equation}
\vspace{-3.5em}
\begin{equation}
PAP = \sum_{i \in AP} \left( \mu_i \Lambda + \Theta_i IR \right)
\label{eq:ap_power}
\end{equation}
\vspace{-2.5em}
\begin{equation}
PCP = \sum_{i \in CP} \left( \sigma_i \Lambda + \Theta_i IR \right)
\label{eq:cp_power}
\end{equation}
\vspace{-2.5em}
\begin{equation}
PQ = \sum_{i \in Q} \left( \lambda_i \phi_{q_i} + \Theta_i \gamma_i \right)
\label{eq:network_device_power}
\end{equation}
\vspace{-2.5em}
\begin{equation}
AR = \sum_{i \in MR} \left( \mu_i \Lambda + \Theta_i IR \right)
\label{eq:AR}
\end{equation}
\vspace{-2.5em}
\begin{equation}
AY = \sum_{i \in MY} \left( \mu_i \Delta + \Theta_i IY \right)
\label{eq:AY}
\end{equation}
\vspace{-2.5em}
\begin{equation}
AGB = \sum_{i \in MGB} \left( \mu_i \Pi + \Theta_i IGB \right)
\label{eq:AGB}
\end{equation}
\vspace{-2.5em}
\begin{equation}
PR = \sum_{i \in DR} \left( \sigma_i \Lambda + \Theta_i IR \right)
\label{eq:PR}
\end{equation}
\vspace{-2.5em}
\begin{equation}
PY = \sum_{i \in DY} \left( \sigma_i \Delta + \Theta_i IY \right)
\label{eq:PY}
\end{equation}
\vspace{-2.5em}
\begin{equation}
PGB = \sum_{i \in DGB} \left( \sigma_i \Pi + \Theta_i IGB \right)
\label{eq:PGB}
\end{equation}
\vspace{-2.5em}
\begin{equation}
PONU = \sum_{i \in U} \left( \sigma_i \chi + \Theta_i \mathbb{U} \right)
\label{eq:PONU}
\end{equation}
\vspace{-2.5em}

\vspace{-0.5em}
The model is subject to the following constraints:

\textbf{Processing Assignment Constraints:}
\vspace{-1.5em}
\begin{align}
\sum_{d \in P} \psi^{ud} &= D_u, \quad \forall u \in UD \label{eq:demand_assignment} \\
\sum_{u \in UD} \psi^{ud} &\leq \Omega_d, \quad \forall d \in P \label{eq:capacity_constraint} \\
\sum_{d \in P} \xi^{ud} &= 1, \quad \forall u \in UD \label{eq:single_assignment}
\end{align}

\vspace{-0.5em}
Constraint~\eqref{eq:demand_assignment} ensures that the total processing demand of user device $u$ is fully assigned across all processing nodes. 

Constraint~\eqref{eq:capacity_constraint} ensures that the total processing demands allocated to any processing node $d$ do not exceed its available capacity $\Omega_d$.

Constraint~\eqref{eq:single_assignment} ensures that each user device's processing demand can be assigned to a single processing node.

\textbf{Traffic Constraints:}
\vspace{-1.5em}
\begin{align}
\lambda^{ud} &= DRR_u \psi^{ud}, \quad \forall u \in UD, d \in P \label{eq:traffic_demand}
\end{align}

\vspace{-1.5em}
Constraint~\eqref{eq:traffic_demand} ensures that the traffic flow between a source and destination pair is equal to the total traffic demand.

\vspace{-1.5em}
\begin{equation}
\sum_{n \in N_m, m \neq n} \lambda_{mn}^{ud} - \sum_{n \in N_m, m \neq n} \lambda_{nm}^{ud} = 
\begin{cases}
\lambda^{ud} & \text{if } m = u \\
-\lambda^{ud} & \text{if } m = d \\
0 & \text{otherwise}
\end{cases}, \quad \forall u \in UD, d \in P, m \in N
\label{eq:flow_conservation}
\end{equation}

\vspace{-0.5em}
Constraint~\eqref{eq:flow_conservation} enforces flow conservation in the P2P-PON network by ensuring that traffic leaving the source equals traffic arriving at the destination, while all intermediate nodes maintain zero net flow.

\vspace{-1.5em}
\begin{equation}
\sum_{u \in UD} \sum_{d \in P} \lambda_{mn}^{ud} \leq L_{mn}, \quad \forall m \in N, n \in N_m
\label{eq:link_capacity}
\end{equation}

\vspace{-0.5em}
Constraint~\eqref{eq:link_capacity} ensures that the total traffic flow on any physical link $(m,n)$ does not exceed the link's capacity $L_{mn}$. 

% ========================================
% BINARY ACTIVATION CONSTRAINTS
% ========================================

\vspace{-2.5em}
\begin{equation}
Z\xi^{ud} \geq \psi^{ud}, \quad \forall u \in UD, d \in P
\label{eq:xi_psi_1}
\end{equation}
\vspace{-3em}
\begin{equation}
\xi^{ud} \leq Z \psi^{ud}, \quad \forall u \in UD, d \in P
\label{eq:xi_psi_2}
\end{equation}

\vspace{-1.5em}
Constraints~\eqref{eq:xi_psi_1} and~\eqref{eq:xi_psi_2} link the binary assignment variable $\xi^{ud}$ to the continuous processing demand $\psi^{ud}$. These constraints ensure that $\xi^{ud} = 1$ when processing demand is assigned from user $u$ to processing node $d$ (i.e., when $\psi^{ud} > 0$), and $\xi^{ud} = 0$ otherwise.

\vspace{-2em}
\begin{equation}
Z \beta_d \geq \sum_{u \in UD} \xi^{ud}, \quad \forall d \in P
\label{eq:beta_xi_1}
\end{equation}
\vspace{-2.5em}
\begin{equation}
\beta_d \leq Z \sum_{u \in UD} \xi^{ud}, \quad \forall d \in P
\label{eq:beta_xi_2}
\end{equation}

\vspace{-0.5em}
Constraints~\eqref{eq:beta_xi_1} and~\eqref{eq:beta_xi_2} ensure that processing node $d$ is activated (i.e., $\beta_d = 1$) when at least one user assigns their processing demand to it. This is essential for correctly calculating the idle power consumption $I_d$ of processing nodes in the total power consumption.

\vspace{-2em}
\begin{equation}
Z \Theta_i \geq \lambda_i, \quad \forall i \in Q
\label{eq:theta_lambda_1}
\end{equation}
\vspace{-3em}
\begin{equation}
\Theta_i \leq Z\lambda_i, \quad \forall i \in Q
\label{eq:theta_lambda_2}
\end{equation}

\vspace{-1.5em}
Constraints~\eqref{eq:theta_lambda_1} and~\eqref{eq:theta_lambda_2} ensure that networking device $i$ is activated (i.e., $\Theta_i = 1$) when traffic flows through it. This is necessary for calculating the idle power consumption $\gamma_i$ of networking devices such as the OLT, switches, and routers.

\vspace{-2.5em}
\begin{equation}
Z \Theta_i \geq \sigma_i, \quad \forall i \in CP \cup PUD
\label{eq:theta_sigma_1}
\end{equation}
\vspace{-3em}
\begin{equation}
\Theta_i \leq Z \sigma_i, \quad \forall i \in CP \cup PUD
\label{eq:theta_sigma_2}
\end{equation}

\vspace{-1.5em}
Constraints~\eqref{eq:theta_sigma_1} and~\eqref{eq:theta_sigma_2} ensure that access point $i$ connected to processing user devices is activated (i.e., $\Theta_i = 1$) when traffic is received at that access point. This ensures proper calculation of idle power consumption for access points serving as destinations for processing tasks.

\vspace{-2.5em}
\begin{equation}
Z \Theta_i \geq \mu_i, \quad \forall i \in AP \cup UD
\label{eq:theta_mu_1}
\end{equation}
\vspace{-3em}
\begin{equation}
\Theta_i \leq Z \mu_i, \quad \forall i \in AP \cup UD
\label{eq:theta_mu_2}
\end{equation}

\vspace{-1.5em}
Constraints~\eqref{eq:theta_mu_1} and~\eqref{eq:theta_mu_2} ensure that access point $i$ connected to source user devices is activated (i.e., $\Theta_i = 1$) when traffic is generated from that access point. This ensures proper calculation of idle power consumption for access points serving user devices that generate processing demands.

\textbf{Delay constraints:}
\vspace{-1.5em}
% Total queuing delay calculation
\begin{equation}
LD_{ud} = \sum_{m \in N} \sum_{n \in N_m, m \neq n} X_{mn}^{ud}, \quad \forall u \in UD, d \in P, u \neq d
\label{eq:total_queuing_delay}
\end{equation}

\vspace{-1em}
Equation~\eqref{eq:total_queuing_delay} calculates the total queuing delay for traffic sent from source node $u$ to destination node $d$ by summing the queuing delays experienced on all links traversed.

\vspace{-2em}
% Total delay for all users
\begin{equation}
TD = \sum_{u \in UD} ED_u
\label{eq:total_delay_all_users}
\end{equation}

\vspace{-1em}
Equation~\eqref{eq:total_delay_all_users} computes the total delay experienced by all users requesting task processing.

% ========================================
% DELAY CONSTRAINTS
% ========================================

\vspace{-2em}
\begin{equation}
\sum_{u \in UD} \sum_{d \in P, u \neq d} \lambda_{mn}^{ud} = TR_{mn}, \quad \forall m \in N, n \in N_m, m \neq n
\label{eq:link_traffic_aggregation}
\end{equation}

\vspace{-1em}
Constraint~\eqref{eq:link_traffic_aggregation} calculates the total traffic between all source-destination pairs on link $(m,n)$.

\vspace{-2.5em}
\begin{align}
Z \, \, \lambda_{mn}^{ud} &\geq TB_{mn}^{ud}, \quad \forall u \in UD, d \in P, m \in N, n \in N_m, m \neq n, u \neq d \label{eq:traffic_binary_lb} \\
\lambda_{mn}^{ud} &\leq Z \quad TB_{mn}^{ud}, \, \, \forall u \in UD, d \in P, m \in N, n \in N_m, m \neq n, u \neq d \label{eq:traffic_binary_ub}
\end{align}

\vspace{-1em}
Constraints~\eqref{eq:traffic_binary_lb} and \eqref{eq:traffic_binary_ub} provide the binary equivalent of $\lambda_{mn}^{ud}$, where $TB_{mn}^{ud} = 1$ if there is traffic between source-destination pair $(u,d)$ on link $(m,n)$, and $TB_{mn}^{ud} = 0$ otherwise.

\vspace{-2em}
\begin{equation}
X_{mn}^{ud} = H_{mn} TB_{mn}^{ud}, \quad \forall u \in UD, d \in P, m \in N, n \in N_m, m \neq n, u \neq d
\label{eq:queuing_delay_nonlinear}
\end{equation}
\vspace{-4em}
\begin{align}
X_{mn}^{ud} &\leq UB_{mn} TB_{mn}^{ud}, \quad \forall u \in UD, d \in P, m \in N, n \in N_m, m \neq n, u \neq d \label{eq:queuing_delay_lin1} \\
X_{mn}^{ud} &\leq H_{mn}, \quad \forall u \in UD, d \in P, m \in N, n \in N_m, m \neq n, u \neq d \label{eq:queuing_delay_lin2} \\
X_{mn}^{ud} &\geq H_{mn} - UB_{mn} (1 - TB_{mn}^{ud}), \quad \forall u \in UD, d \in P, m \in N, n \in N_m, m \neq n, u \neq d \label{eq:queuing_delay_lin3}
\end{align}

\vspace{-1em}
Constraints~\eqref{eq:queuing_delay_lin1}--\eqref{eq:queuing_delay_lin3} linearize the nonlinear equation~\eqref{eq:queuing_delay_nonlinear}. These constraints calculate the queuing delay experienced by traffic flow from source $u$ to destination $d$ traversing link $(m,n)$.

\vspace{-3em}
\begin{align}
Z \, \, TR_{mn} &\geq HB_{mn}, \quad \forall m \in N, n \in N_m, m \neq n \label{eq:link_utilization_lb} \\
TR_{mn} &\leq Z \, \, HB_{mn}, \quad \forall m \in N, n \in N_m, m \neq n \label{eq:link_utilization_ub}
\end{align}

\vspace{-1em}
Constraints~\eqref{eq:link_utilization_lb} and \eqref{eq:link_utilization_ub} provide the binary equivalent of $TR_{mn}$, where $HB_{mn} = 1$ if there is traffic passing through link $(m,n)$, and $HB_{mn} = 0$ otherwise.

\vspace{-2em}
\begin{equation}
H_{mn} \geq RA_{mnp} TR_{mn} + TI_{mnp}, \quad \forall m \in N, n \in N_m, p \in PI, m \neq n
\label{eq:piecewise_delay}
\end{equation}

Constraint~\eqref{eq:piecewise_delay} calculates the queuing delay experienced on link $(m,n)$. The calculation provides a linear approximation based on piecewise linearization of the queuing delay function.

\vspace{-2.5em}
\begin{equation}
ED_u = \beta_d \, \, LD_{ud}, \quad \forall u \in UD, d \in P, u \neq d
\label{eq:e2e_delay_nonlinear}
\end{equation}
\vspace{-3em}
\begin{equation}
\sum_{u \in UD} ED_u \leq Z \sum_{d \in P} \beta_d \label{eq:e2e_delay_lin1}
\end{equation}
\vspace{-2em}
\begin{equation}
ED_u \leq \sum_{d \in P, u \neq d} LD_{ud}, \quad \forall u \in UD \label{eq:e2e_delay_lin2}
\end{equation}
\vspace{-2em}
\begin{equation}
ED_u \geq LD_{ud} - (1 - \beta_d) \, \, Z, \quad \forall u \in UD, d \in P, u \neq d \label{eq:e2e_delay_lin3}
\end{equation}

\vspace{-1em}
Constraints~\eqref{eq:e2e_delay_lin1}--\eqref{eq:e2e_delay_lin3} linearize Equation~\eqref{eq:e2e_delay_nonlinear}, which involves a binary variable multiplied by a continuous variable. These constraints associate the delay calculation with activated processing nodes, ensuring that delay is only calculated for traffic between active source-destination pairs $(u,d)$.

\section{Performance Evaluation}
\label{sec:per_eval}

This section presents the performance evaluation of the proposed P2P-PON architecture in comparison with the AWGR-PON architecture. The evaluation focuses on two key aspects: (i) energy efficiency, in terms of processing, networking, and total power consumption, and (ii) network performance, assessed through end-to-end queuing delay. Multiple scenarios are considered to capture the impact of user distribution and processing demand variations within the building. The results highlight the trade-offs between power consumption and network delay under different workload conditions. The objective of this evaluation is to quantify how architectural connectivity constraints influence energy consumption, delay behavior, and workload placement decisions under varying traffic locality and intensity.

The evaluation employs the multi-objective optimization framework described in Section~\ref{subsec:proposed}, which jointly considers total power consumption and end-to-end queuing delay. Two optimization modes are examined: power-aware optimization, where power consumption is prioritized, and delay-aware optimization, where queuing delay is explicitly minimized alongside power consumption by adjusting the weighting coefficients $\alpha$ and $\beta$. Due to the computational complexity introduced by the queuing delay constraints, delay-aware evaluation is limited to a representative low-intensity scenario, while the remaining scenarios focus on power-aware optimization. The delay performance of the architectures is therefore analyzed separately in Section~V-B to highlight fundamental architectural differences.

\subsection{Simulation Parameters}

This section summarizes the system parameters and assumptions used in the performance evaluation. The proposed P2P-PON architecture and the AWGR-PON baseline are both evaluated under identical networking and processing configurations to ensure a fair comparison. The parameters of the networking devices and processing nodes are listed in Tables~\ref{tab:networking_params} and~\ref{tab:processing_params}, respectively \cite{fadlelmula2025energy}.

The same processing capacities, idle and maximum power consumption values, and energy efficiencies are assumed for corresponding devices in both architectures. The only exception is the optical network unit (ONU), where the P2P-PON employs fixed ONUs with a power consumption of 10~W, while the AWGR-PON relies on tunable ONUs with higher power consumption \cite{HuaweiOptiXstarP813E_E}. The backplane in the P2P-PON architecture is modeled as a passive component and therefore does not contribute to power consumption.

Processing demands generated by user devices range from 6 to 20~GFLOPs, while the associated traffic demands vary between 0.3 and 1~Gbps. Multiple demand distributions are considered to capture different levels of traffic locality, including scenarios where demands are generated by multiple users across different rooms as well as highly localized scenarios where all demands originate within a single room.

\begin{table*}[htbp]
\centering
\footnotesize
\caption{Networking Devices Parameters}
\label{tab:networking_params}
\begin{tabular}{>{\centering\arraybackslash}p{3cm} 
                >{\centering\arraybackslash}p{3cm} 
                >{\centering\arraybackslash}c 
                >{\centering\arraybackslash}c 
                >{\centering\arraybackslash}c 
                >{\centering\arraybackslash}c}
\hline
\makecell{\textbf{Network} \\ \textbf{Device}} & 
\makecell{\textbf{Type /} \\ \textbf{Wavelength}} & 
\makecell{\textbf{Maximum} \\ \textbf{Power (W)}} & 
\makecell{\textbf{Idle} \\ \textbf{Power (W)}} & 
\makecell{\textbf{Capacity} \\ \textbf{(GFLOPs)}} & 
\makecell{\textbf{Efficiency} \\ \textbf{(W/GFLOPs)}} \\
\hline
Access Point (AP) & Red & 7.2 & 4.32 & 2.5 & 1.52 \\
                  & Yellow & 4.5 & 2.7 & 2.5 & 0.72 \\
                  & Green & 2.7 & 1.62 & 2.5 & 0.432 \\
                  & Blue & 2.7 & 1.62 & 2.25 & 0.485 \\
ONU               & - & 10 & 6 & 10 & 0.9 \\
Ethernet Switch   & - & 300 & 180 & 160 & 1.125 \\
Aggregation Switch & - & 435 & 261 & 240 & 0.725 \\
Edge Router       & - & 435 & 261 & 240 & 0.725 \\
Optical Switch    & - & 750 & 450 & 480 & 0.625 \\
Core Router       & - & 344 & 206.4 & 3200 & 0.043 \\
\hline
\end{tabular}
\end{table*}

\begin{table*}[htbp]
\centering
\footnotesize
\caption{Processing Devices Parameters}
\label{tab:processing_params}

\begin{tabular}{>{\centering\arraybackslash}p{3cm} 
                >{\centering\arraybackslash}c 
                >{\centering\arraybackslash}c 
                >{\centering\arraybackslash}c 
                >{\centering\arraybackslash}c}
\hline
\makecell{\textbf{Processing} \\ \textbf{Node}} & 
\makecell{\textbf{Maximum} \\ \textbf{Power (W)}} & 
\makecell{\textbf{Idle} \\ \textbf{Power (W)}} & 
\makecell{\textbf{Capacity} \\ \textbf{(GFLOPs)}} & 
\makecell{\textbf{Efficiency} \\ \textbf{(W/GFLOPs)}} \\
\hline
Cloud Server & 1100 & 660 & 1612.8 & 0.27 \\
MFS          & 750  & 450 & 403.2  & 0.74 \\
CFS          & 350  & 210 & 121.6  & 1.15 \\
BFS          & 305  & 183 & 99     & 1.23 \\
RFS          & 65   & 39  & 64     & 0.41 \\
User Devices & 18   & 10.8 & 12.288 & 0.55 \\
\hline
\end{tabular}
\end{table*}

\subsection{Processing Placement and Power Consumption Evaluation}

This subsection evaluates the proposed P2P-PON architecture in comparison with the AWGR-PON architecture under power-aware optimization. The analysis focuses on optimal processing placement decisions and their impact on processing, networking, and total power consumption.

\subsubsection{Scenario 1: Two Source User Devices per Room}

\begin{figure}[h]
\centering
\includegraphics[width=0.7\textwidth]{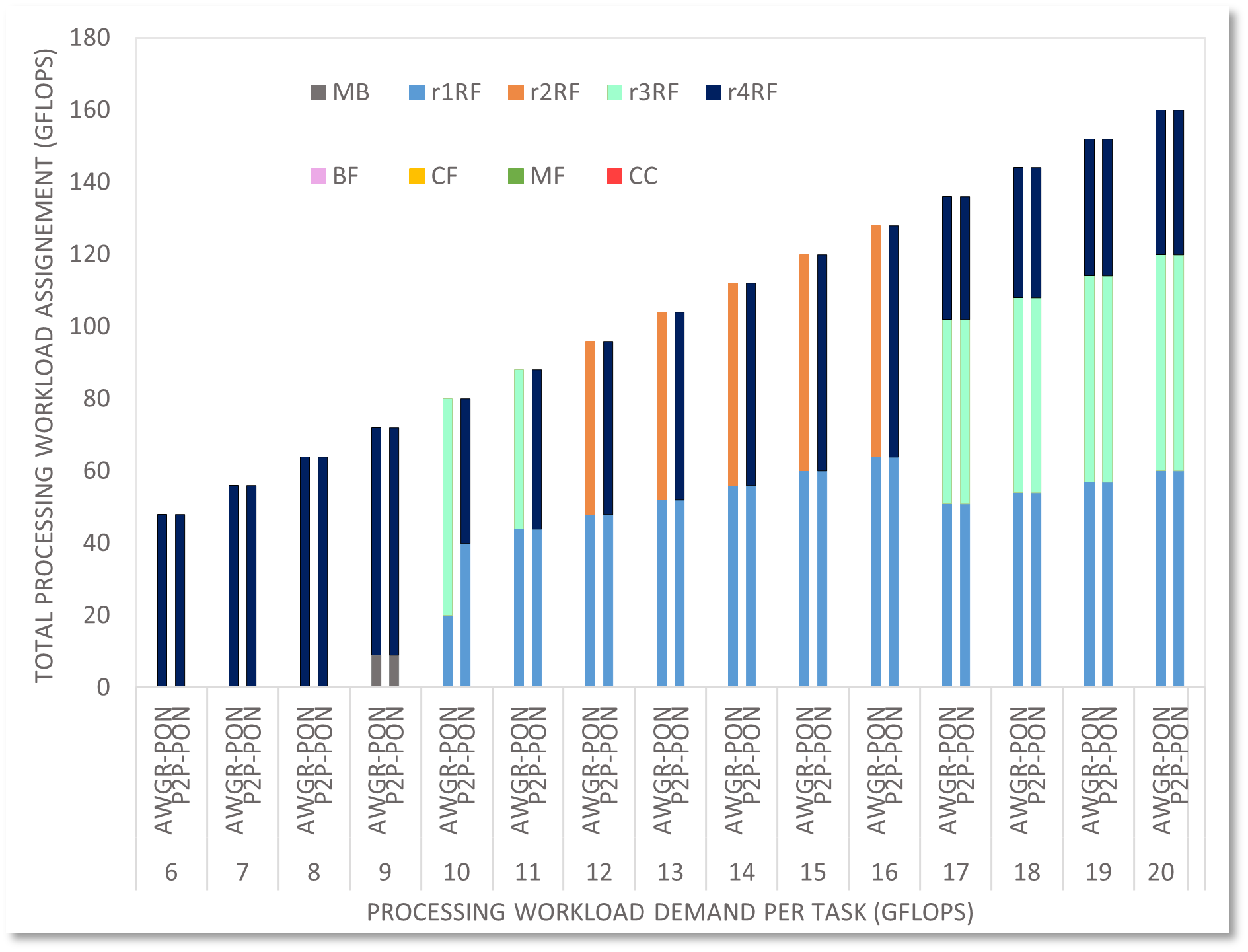} % scales image to 60% of text width
\caption{Optimal processing placement in the AWGR-PON and P2P-PON for the first scenario.}
\label{figure2}
\end{figure}

This scenario examines the allocation of processing and networking demands when processing demands originate from two user devices in each room. Fig.~\ref{figure2} illustrates the optimal processing placement obtained for the P2P-PON architecture in comparison with the AWGR-PON architecture. The results indicate identical processing placement decisions for both architectures. This outcome is primarily attributed to the passive nature of the networking components in both architectures, which leads to equivalent processing consolidation behavior.

\begin{figure}[h]
    \centering
    \begin{subfigure}[b]{0.32\textwidth}
        \centering
        \includegraphics[height=4cm, width=\textwidth]{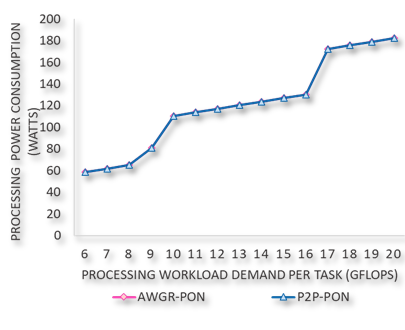}
        \caption{\small Processing power consumption}
        \label{fig:processing_power1}
    \end{subfigure}
    \hfill
    \begin{subfigure}[b]{0.32\textwidth}
        \centering
        \includegraphics[height=4cm, width=\textwidth]{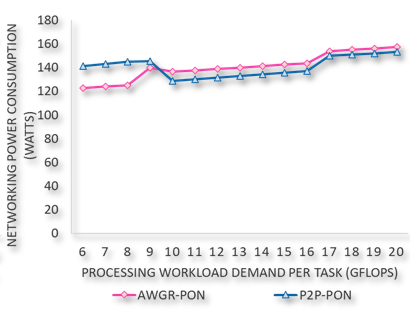}
        \caption{\small Networking power consumption}
        \label{fig:networking_power1}
    \end{subfigure}
    \hfill
    \begin{subfigure}[b]{0.32\textwidth}
        \centering
        \includegraphics[height=4cm, width=\textwidth]{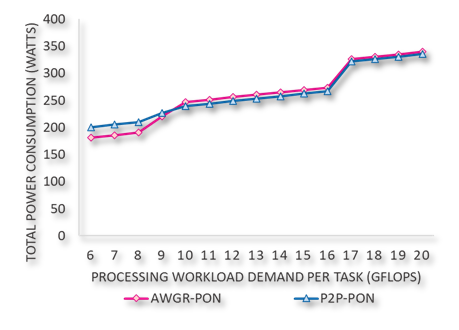}
        \caption{\small Total power consumption}
        \label{fig:total_power1}
    \end{subfigure}
    \caption{Power consumption comparison between the AWGR-PON and the P2P-PON for the first scenario.}
    \label{fig:pon2_comparison}
\end{figure}

Although the processing placement results are identical, differences arise in the power consumption performance. 
Fig.~\ref{fig:pon2_comparison} presents the processing, networking, and total power consumption for both architectures. 
Due to the identical allocation decisions, the processing power consumption of the two architectures is exactly the same, as shown in Fig.~\ref{fig:processing_power1}.

At low processing demands (6--8~GFLOPs), the P2P-PON architecture exhibits higher networking power consumption than the AWGR-PON, as illustrated in Fig.~\ref{fig:networking_power1}. In the AWGR-PON, accessing remote RFSs does not introduce additional networking power consumption because all intermediate devices remain passive. 
In contrast, the P2P-PON architecture may require traffic relaying through active access points to reach remote RFSs, resulting in increased networking power consumption. As processing demand increases, the networking power consumption of the P2P-PON decreases and approaches that of the AWGR-PON. At 9~GFLOPs, both architectures exhibit comparable networking power consumption. 
This behavior is attributed to the reuse of already active access points for traffic relaying in the P2P-PON, which limits the activation of additional networking devices. Conversely, the AWGR-PON requires the activation of additional access points to accommodate the increased demand. For processing demands beyond 10~GFLOPs, multiple room fog servers become active, enabling more localized processing and reducing the need for long relay paths in the P2P-PON architecture. As a result, the P2P-PON achieves lower networking power consumption, yielding an average reduction of approximately 6\% compared to the AWGR-PON. 
Furthermore, the absence of tunable ONUs in the P2P-PON contributes to additional networking power savings.

The total power consumption trends, shown in Fig.~\ref{fig:total_power1}, closely follow those of the networking power consumption. Overall, the reduced networking power in the P2P-PON translates into an average total power consumption saving of approximately 3\% relative to the AWGR-PON in this scenario.

\subsubsection{Scenario 2:  Six Source User Devices per Room}
\begin{figure}[h]
\centering
\includegraphics[width=0.7\textwidth]{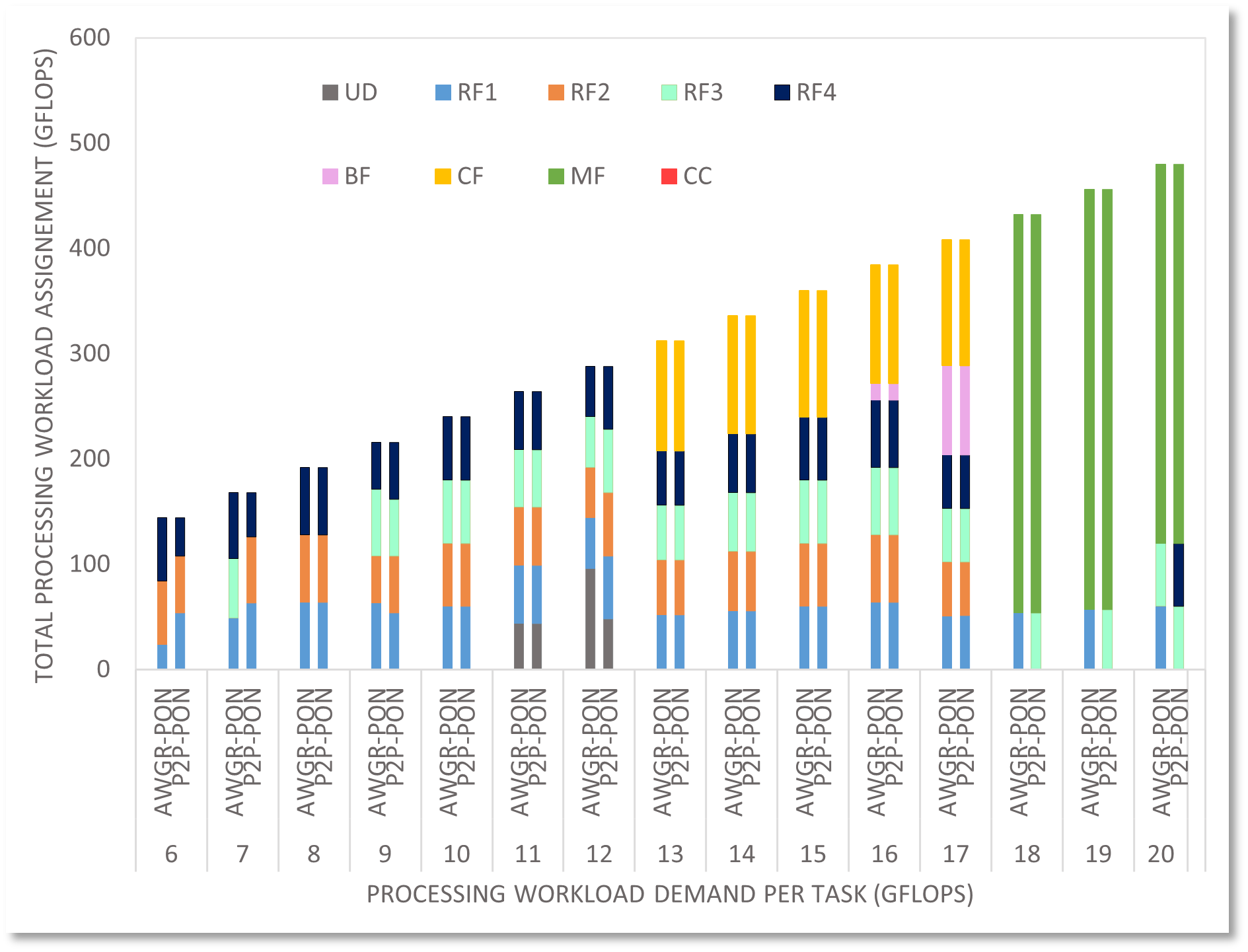} % scales image to 60% of text width
\caption{Optimal processing placement in the AWGR-PON and P2P-PON for the second scenario.}
\label{figure4}
\end{figure}

This scenario evaluates the performance of the two architectures under higher load conditions, where processing demands are generated by six user devices in each room. Fig.~\ref{figure4} shows the optimal processing placement obtained for the AWGR-PON and P2P-PON architectures. Overall, the results exhibit similar processing allocation patterns in both architectures, with only a limited number of differences observed at specific demand levels.

The first noticeable difference occurs at a processing demand of 6~GFLOPs. 
In the AWGR-PON architecture, RFSs are sequentially fully utilized before additional servers are activated. In contrast, the P2P-PON architecture does not strictly follow a full-packing strategy. Instead, processing placement is influenced by the availability of direct connections between access points and RFSs, which may lead to partial utilization of multiple servers to avoid additional traffic relaying. Similar allocation behavior is observed at 7 and 9~GFLOPs, where minor differences in workload placement arise primarily from architectural connectivity constraints rather than processing capacity limitations. A second notable deviation appears at 12~GFLOPs, where the AWGR-PON allocates a larger portion of the workload to user devices compared to the P2P-PON. This behavior is attributed to networking constraints that limit the effective utilization of room fog servers in the AWGR-PON. In contrast, the higher networking flexibility of the P2P-PON enables more efficient utilization of the available room fog servers. For higher processing demands, both architectures exhibit similar processing placement trends.

\begin{figure}[h]
    \centering
    \begin{subfigure}[b]{0.32\textwidth}
        \centering
        \includegraphics[height=4cm, width=\textwidth]{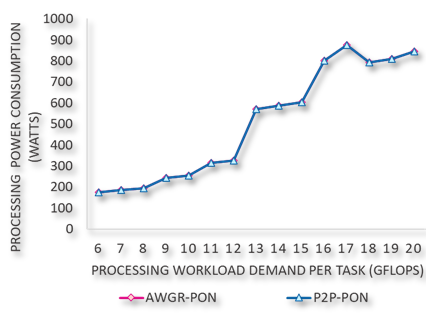}
        \caption{\small Processing power consumption}
        \label{fig:processing_power2}
    \end{subfigure}
    \hfill
    \begin{subfigure}[b]{0.32\textwidth}
        \centering
        \includegraphics[height=4cm, width=\textwidth]{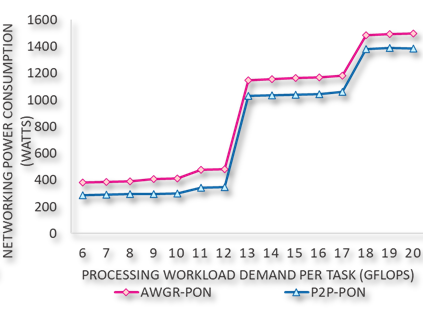}
        \caption{\small Networking power consumption}
        \label{fig:networking_power2}
    \end{subfigure}
    \hfill
    \begin{subfigure}[b]{0.32\textwidth}
        \centering
        \includegraphics[height=4cm, width=\textwidth]{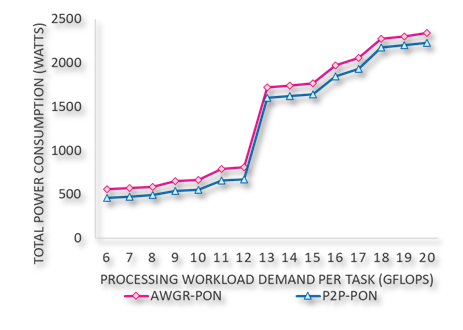}
        \caption{\small Total power consumption}
        \label{fig:total_power2}
    \end{subfigure}
    \caption{Power consumption comparison between the AWGR-PON and the P2P-PON for the second scenario.}
    \label{fig:pon2_comparison2}
\end{figure}

Fig.~\ref{fig:pon2_comparison2} compares the processing, networking, and total power consumption of the two architectures.  As in the previous scenario, identical processing placement decisions lead to identical processing power consumption values, as shown in Fig.~\ref{fig:processing_power2}. In contrast, the P2P-PON consistently achieves lower networking power consumption than the AWGR-PON, as illustrated in Fig.~\ref{fig:networking_power2}. This improvement is primarily attributed to the use of lower-power fixed ONUs in the P2P-PON architecture. Networking power consumption savings of up to 14\% are observed. The reduction in networking power consumption translates directly into total power consumption savings of 8\%, as shown in  Fig.~\ref{fig:total_power2}.

\subsubsection{Scenario 3: Eight Source User Devices in the Same Room}
\begin{figure}[h]
\centering
\includegraphics[width=0.7\textwidth]{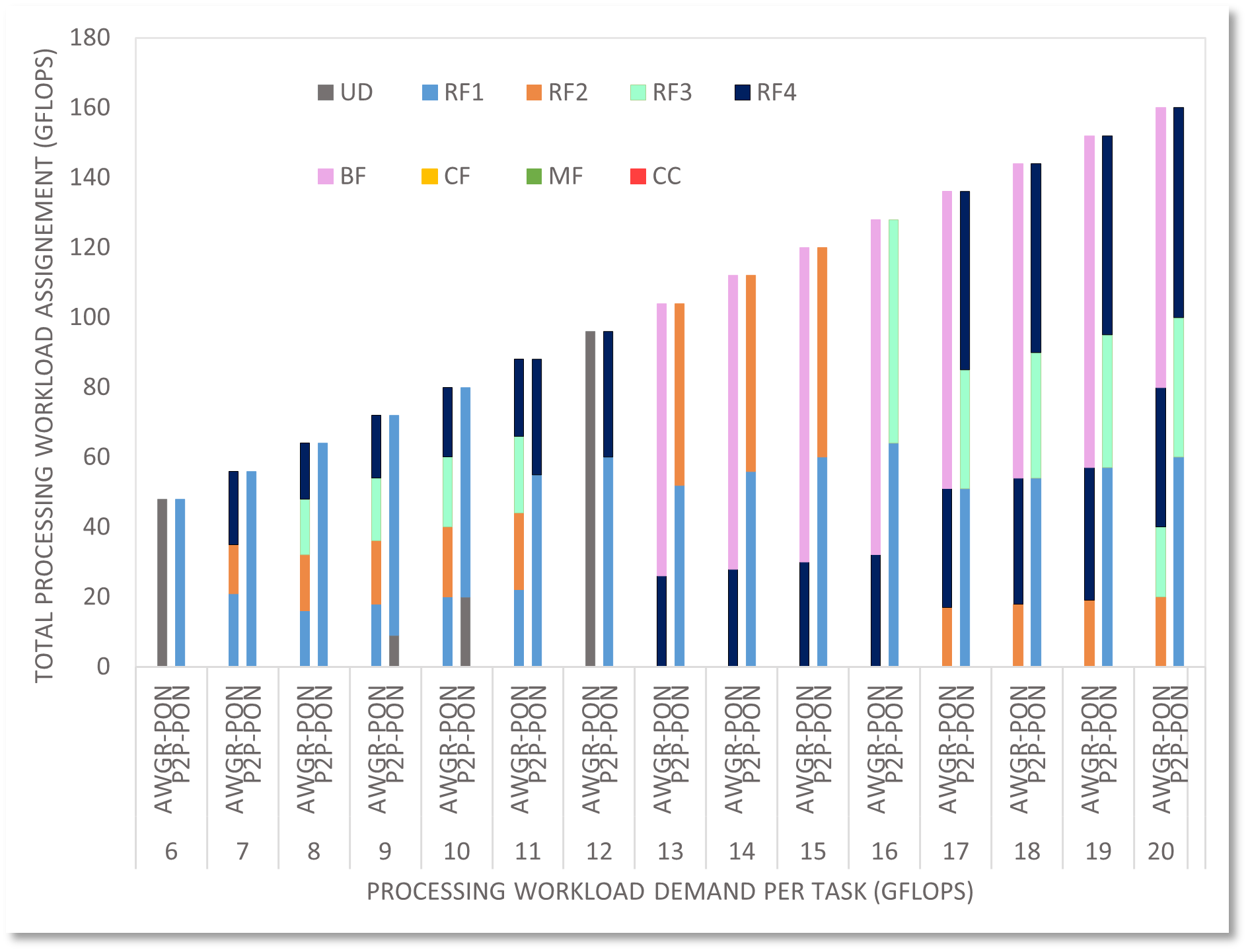} % scales image to 60% of text width
\caption{Optimal processing placement in the AWGR-PON and P2P-PON for the third scenario.}
\label{figure6}
\end{figure}

This scenario evaluates the behavior of the two architectures when all processing demands originate from user devices located within a single room. 
Fig.~\ref{figure6} shows the resulting optimal processing placement for the AWGR-PON and P2P-PON architectures. This scenario clearly exposes the impact of networking capacity limitations in the AWGR-PON architecture and their influence on workload allocation decisions. 

At lower processing demands (6--8~GFLOPs), the P2P-PON consolidates all workloads into a RFS. 
Although this consolidation incurs additional networking power consumption due to the activation of relaying access points, it remains more energy efficient than activating multiple processing nodes. 
In contrast, the AWGR-PON is unable to achieve similar consolidation as it activates user devices instead of a single RFS due to networking bottlenecks that restrict effective utilization of the RFSs. 
This limitation results in the activation of three and four RFSs at 7 and 8-11~GFLOPs, respectively. In the P2P-PON architecture, user devices assist the RFS at 9 and 10~GFLOPs; however, at 11~GFLOPs, activating an additional RFS becomes more energy efficient than further reliance on user devices. From 12 to 16~GFLOPs, two RFSs are sufficient to host the demands in the P2P-PON, while at 17~GFLOPs an additional RFS is activated to support the remaining workload up to 20~GFLOPs. In contrast, the allocation behavior of the AWGR-PON differs significantly at higher demand levels. At 12~GFLOPs, the availability of networking capacity toward higher-tier nodes leads the model to assign all demands to user devices to avoid activating additional processing nodes. From 13~GFLOPs onward, the processing capacity of user devices is exhausted, necessitating the activation of the BFS in conjunction with an RFS. At 17~GFLOPs, an additional RFS is activated, and by 20~GFLOPs the workload must be distributed across three RFSs in addition to previously activated nodes.

\begin{figure}[h]
    \centering
    \begin{subfigure}[b]{0.32\textwidth}
        \centering
        \includegraphics[height=4cm, width=\textwidth]{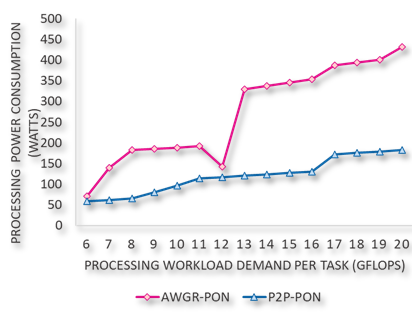}
        \caption{\small Processing power consumption}
        \label{fig:processing_power3}
    \end{subfigure}
    \hfill
    \begin{subfigure}[b]{0.32\textwidth}
        \centering
        \includegraphics[height=4cm, width=\textwidth]{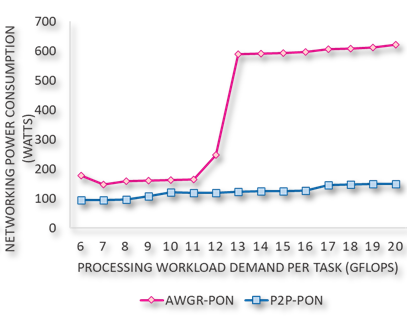}
        \caption{\small Networking power consumption}
        \label{fig:networking_power3}
    \end{subfigure}
    \hfill
    \begin{subfigure}[b]{0.32\textwidth}
        \centering
        \includegraphics[height=4cm, width=\textwidth]{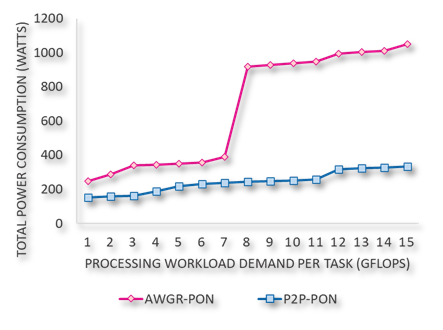}
        \caption{\small Total power consumption}
        \label{fig:total_power3}
    \end{subfigure}
    \caption{Power consumption comparison between the AWGR-PON and the P2P-PON for the third scenario.}
    \label{fig:pon2_comparison3}
\end{figure}

The corresponding power consumption results are shown in Fig.~\ref{fig:pon2_comparison3}. Due to the substantial differences in processing placement, significant disparities in power consumption are observed between the two architectures. 
As illustrated in Fig.~\ref{fig:processing_power3}, the P2P-PON achieves processing power consumption savings of up to 56\% by activating fewer processing nodes. These savings are primarily enabled by the ability of the P2P-PON to overcome the networking constraints that limit consolidation in the AWGR-PON.

Fig.~\ref{fig:networking_power3} presents the networking power consumption. 
The availability of sufficient passive connectivity, combined with the use of energy-efficient ONUs, enables the P2P-PON to significantly outperform the AWGR-PON. The difference becomes particularly clear beyond 13~GFLOPs, where the AWGR-PON is forced to offload processing demands outside the building, resulting in a sharp increase in networking power consumption. Consequently, networking power savings of up to 69\% are achieved.

The combined processing and networking savings translate into total power consumption reductions of up to 64\%, as shown in Fig.~\ref{fig:total_power3}. 
These results highlight the effectiveness of the P2P-PON architecture in supporting highly localized and concentrated traffic demands under stringent networking constraints.

\begin{figure}[h]
    \centering
    \includegraphics[width=0.5\linewidth]{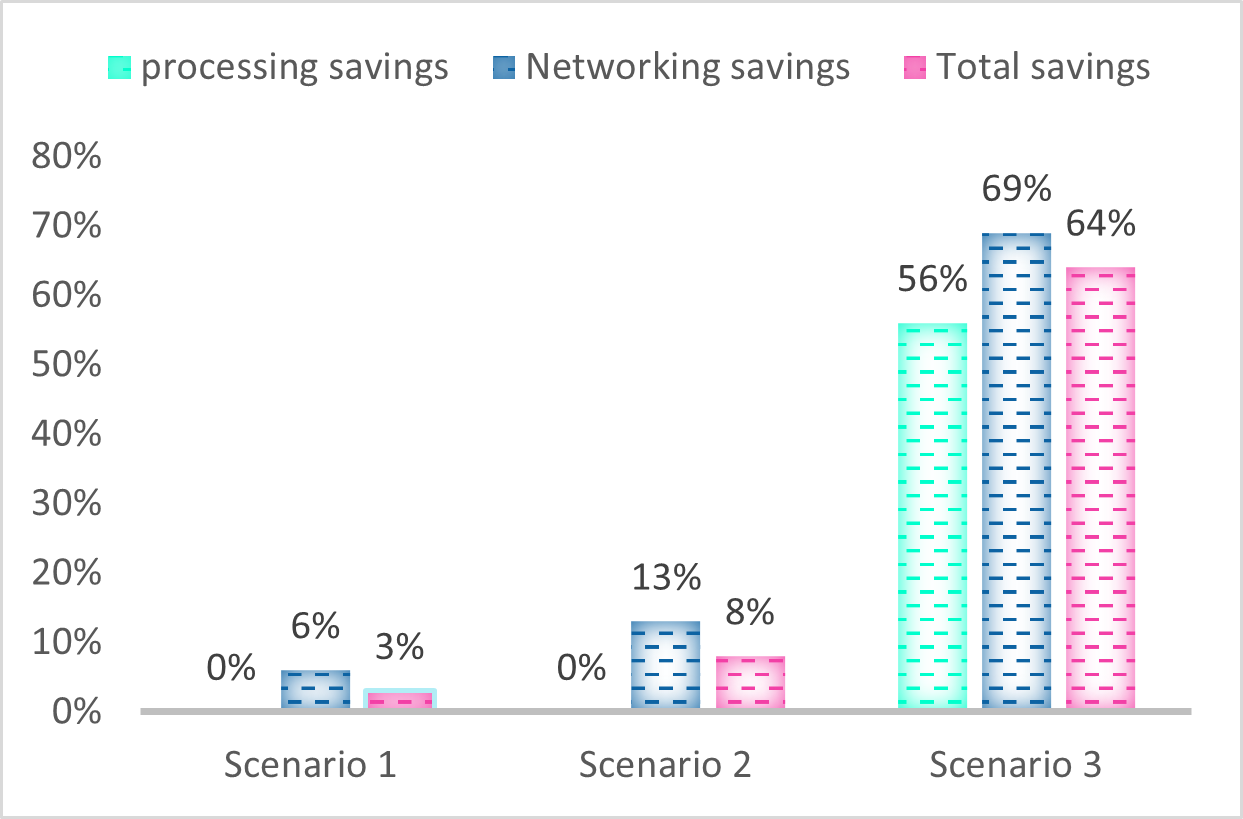}
    \caption{Average processing, networking, and total power consumption savings of the P2P-PON relative to the AWGR-PON across all evaluated scenarios.}
    \label{fig:average_power_savings}
\end{figure}

Figure~\ref{fig:average_power_savings} summarizes the average processing, networking, and total power consumption savings achieved by the P2P-PON architecture relative to the AWGR-PON across all evaluated scenarios.
The results show that the dominant contribution to total power savings originates from reductions in networking power consumption, which become increasingly pronounced in scenarios with higher traffic locality. Processing power savings are more scenario-dependent and are primarily observed when architectural connectivity constraints limit processing consolidation in the AWGR-PON.

\subsection{Queuing Delay Evaluation}

In addition to energy efficiency, latency performance is a key requirement for emerging delay-sensitive applications. In this subsection, the delay performance of the proposed P2P-PON architecture is evaluated and compared against the AWGR-PON architecture in terms of end-to-end queuing delay.

In general, network delay consists of transmission, propagation, processing, and queuing delays \cite{roy2021overview}. Transmission and propagation delays are determined by packet size, link data rate, and physical distance, while processing delay depends on hardware-specific characteristics of network and processing nodes. For a given physical deployment, these delay components are largely traffic-independent and remain unchanged. In contrast, queuing delay is highly sensitive to traffic load and workload placement decisions and therefore represents the dominant and most variable delay component under moderate and high utilization conditions \cite{kurose2019computer}. Accordingly, this study focuses exclusively on queuing delay, while other delay components are assumed constant and are excluded from the optimization.

Queuing delay on network links is modeled using M/M/1 queuing system, where packet arrivals follow a Poisson process with arrival rate $\lambda$ and service times are exponentially distributed with service rate $\mu$. In this work, the arrival rate corresponds to the aggregate traffic load on a link, while the service rate corresponds to the link data rate. 
The resulting queuing delay is a nonlinear function of link utilization, defined as $\rho = \lambda / \mu$ \cite{kurose2019computer}.

\begin{figure}[h]
    \centering
    \includegraphics[width=0.5\linewidth]{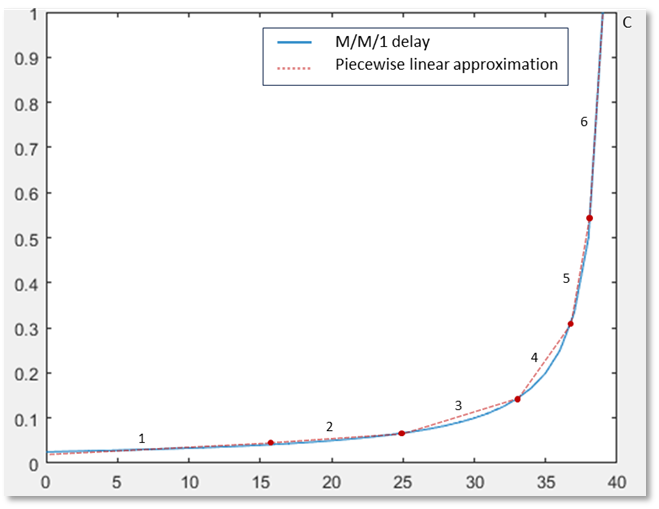}
    \caption{M/M/1 queuing model used for queuing delay estimation.}
    \label{fig:mm1}
\end{figure}

To enable incorporation of queuing delay into the optimization framework, the nonlinear M/M/1 delay function is approximated using a piecewise linear formulation \cite{lin2013review}. Figure~\ref{fig:mm1} illustrates the piecewise linear approximation of the M/M/1 queuing delay employed in this work for a single 40~Gbps link, using six linear segments.

Due to the computational complexity introduced by the queuing delay constraints, the delay-aware evaluation is limited to a representative low-intensity user scenario. This scenario is sufficient to expose fundamental architectural differences in delay behavior between the AWGR-PON and P2P-PON architectures.

\subsubsection{Scenario 1: Two Source User Devices per Room}

In this scenario, two user devices in each room generate processing demands. The scenario is evaluated under both power-aware and delay-aware optimization to illustrate the impact of explicitly accounting for queuing delay on workload placement and performance trade-offs.

\begin{figure}[h]
    \centering
    \includegraphics[width=0.7\linewidth]{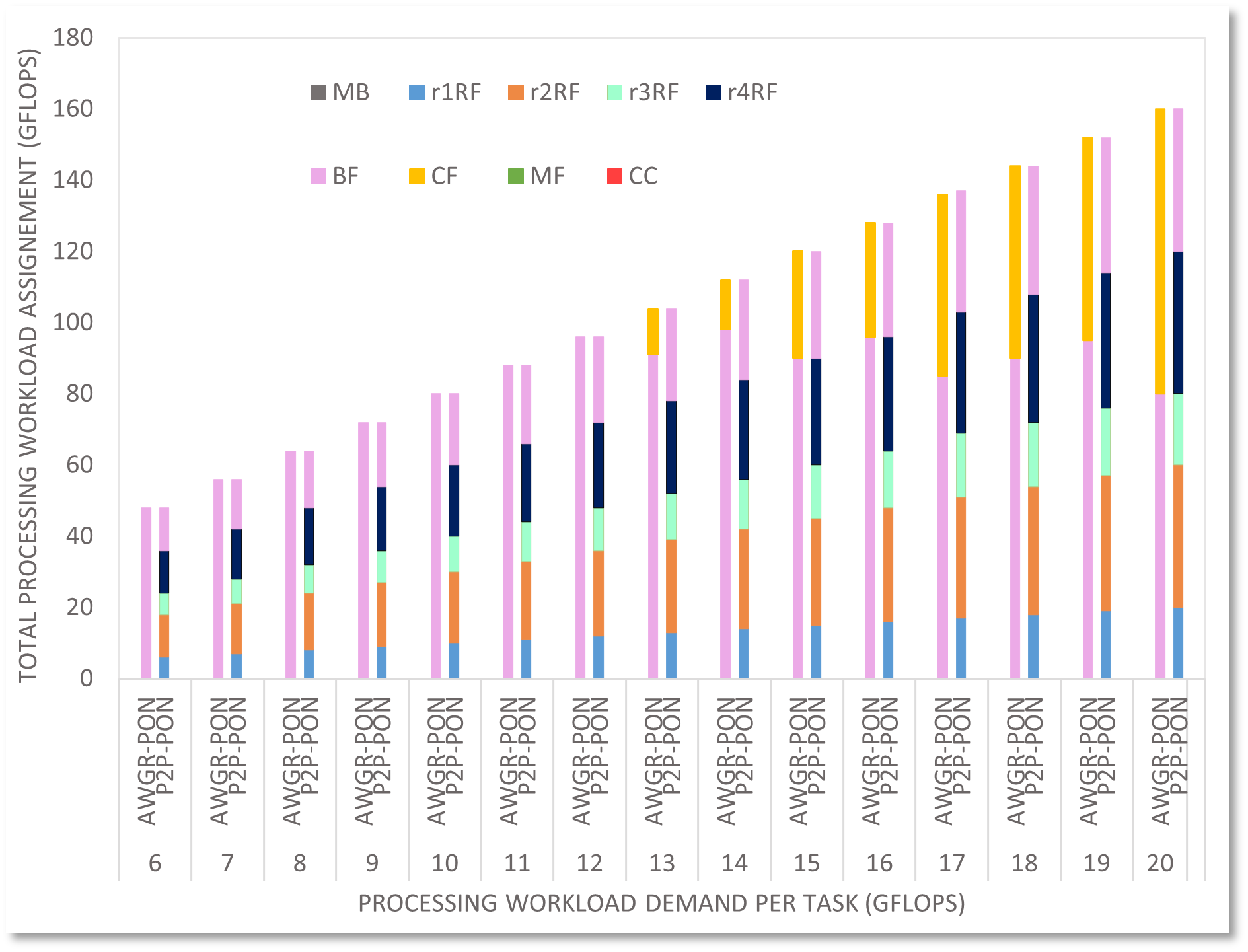}
    \caption{Optimal processing placement under delay minimization for the AWGR-PON and P2P-PON architectures.}
    \label{fig:delay_placement}
\end{figure}

The results obtained under delay-aware optimization, where the power weight is set to a negligible value $(\alpha \ll 1)$, while the delay weight is set to 1 $(\beta = 1)$. The optimal power-minimizing workload placement is illustrated in Fig.~\ref{fig:delay_placement}. For the AWGR-PON architecture, all processing demands are offloaded to the BFS, while in-building processing resources are avoided. This allocation pattern persists until the total demand reaches 13~GFLOPs, at which point the CFS is activated in addition to the BFS. In contrast, the P2P-PON architecture exhibits a more distributed allocation strategy. Part of the workload is processed locally within the building, while the remaining demands are assigned to the BFS. Notably, the model allocates two tasks to the RFSs in rooms 2 and 4, and one task to each of the RFSs in rooms 1 and 3. This behavior is driven by the network topology, where access points in rooms 1 and 3 have direct connectivity to the OLT, making forwarding traffic to the BFS preferable to local processing from a delay perspective.

\begin{figure}[h]
    \centering
    \includegraphics[width=0.7\linewidth]{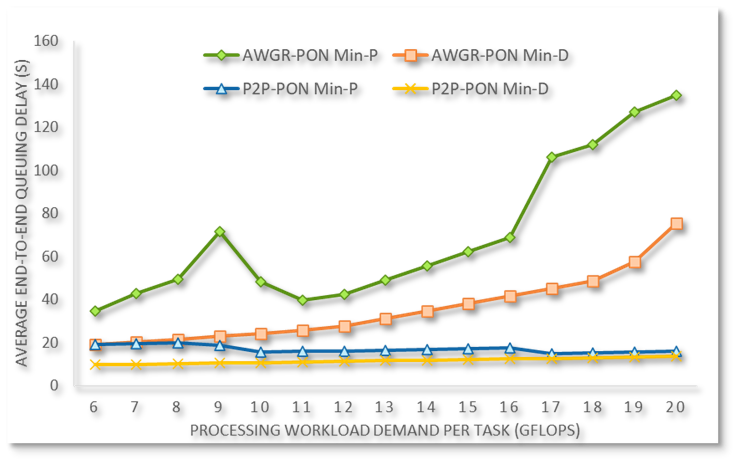}
    \caption{Total end-to-end queuing delay comparison between the AWGR-PON and P2P-PON architectures.}
    \label{fig:queuing_delay}
\end{figure}

The end-to-end queuing delay results for both architectures are shown in Fig.~\ref{fig:queuing_delay}. The reported values represent the total end-to-end queuing delay experienced by all user devices for both power-aware (AWGR-PON-Min-P, P2p-PON-Min-P) and delay-aware (AWGR-PON-Min-D, P2p-PON-Min-D) models. Overall, the P2P-PON architecture consistently achieves lower queuing delay than the AWGR-PON. At lower processing demands, the delay experienced by the AWGR-PON under delay-aware optimization closely matches that of the P2P-PON under power-aware optimization. However, as demand increases, the AWGR-PON exhibits a steady increase in queuing delay. This behavior is primarily attributed to limited link capacities, particularly between user devices and access points. The pronounced delay increase at 9~GFLOPs corresponds to the activation of an additional user device alongside an RFS, which introduces increased congestion on access links. For higher demands, the AWGR-PON under power-aware optimization continues to exhibit increasing delay, whereas the P2P-PON under delay-aware optimization maintains a more stable growth trend. Although some variability is observed under power-aware optimization, the delay gap between the two architectures narrows at higher demand levels (from 17~GFLOPs onward). Overall, the P2P-PON architecture consistently achieves lower average queuing delay than the AWGR-PON. Under power-aware optimization, the P2P-PON reduces average queuing delay by 76\% compared to the AWGR-PON. Under delay-aware optimization, the average delay reduction is 67\%. These results demonstrate that the P2P-PON maintains superior delay performance regardless of the optimization objective.

\begin{figure}[h]
    \centering
    \includegraphics[width=0.7\linewidth]{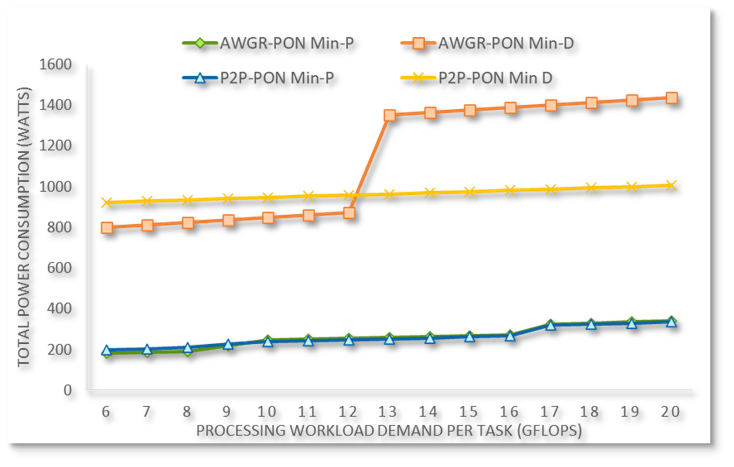}
    \caption{Total power consumption under power-aware and delay-aware optimization.}
    \label{fig:delay_power}
\end{figure}

Fig.~\ref{fig:delay_power} presents the total power consumption results for both architectures under power-aware and delay-aware optimization. When power consumption is minimized, both architectures exhibit comparable performance, with only marginal differences in average power consumption (0.10\% reduction for P2P-PON). However, more pronounced differences emerge under delay-aware optimization.

In the AWGR-PON architecture, all demands between 6 and 12~GFLOPs are allocated to the BFS. Beyond this range, the activation of the CFS in addition to the BFS leads to a substantial increase in total power consumption. In contrast, the P2P-PON distributes workloads across the BFS and RFSs, resulting in a more gradual increase in power consumption. Although the activation of multiple RFSs incurs additional processing power, the overall increase remains modest. From 13~GFLOPs onward, the P2P-PON exhibits significantly lower power consumption than the AWGR-PON, highlighting its superior scalability when delay constraints are considered. Overall, under delay-aware optimization, the P2P-PON achieves a 15.0\% reduction in average power consumption compared to the AWGR-PON, while simultaneously maintaining superior delay performance.

\section{Conclusion}
\label{sec:conc}

This paper proposed a point-to-point passive optical network (P2P-PON) architecture as an energy-efficient and low-latency backhaul solution for VLC-enabled indoor fog computing systems. A mixed integer linear programming (MILP) framework was developed to jointly optimize workload allocation, traffic routing, power consumption, and end-to-end queuing delay across a multi-layer fog computing infrastructure. Performance evaluation demonstrates that the P2P-PON architecture achieves significant improvements over an AWGR-PON-based architecture. Under power-aware optimization, the P2P-PON reduces total power consumption by up to 64\% and average queuing delay by up to 76\%. Under delay-aware optimization, it reduces average queuing delay by 67\% and average power consumption by 15\%. These improvements stem from enhanced in-building connectivity, elimination of wavelength routing constraints, and more effective utilization of distributed fog resources. Future work will extend the framework to handle dynamic traffic, support uplink via hybrid VLC/RF or infrared links, and explore scalable heuristics and reinforcement learning methods for adaptive near-real-time optimization in large-scale deployments.

\bibliography{references}     

\bibliographystyle{IEEEtran}

\newpage


\vfill

\end{document}